# La meridiana di Egnazio Danti nella Torre dei Venti in Vaticano: un'icona della riforma Gregoriana del calendario

di Costantino Sigismondi

## Introduzione: Egnazio Danti ed i suoi strumenti astronomici

La Torre dei Venti domina l'angolo Sud Ovest del cortile della Pigna (nell'area dei Musei Vaticani), ed è inclusa negli ambienti dell'Archivio Segreto Vaticano. Non è aperta al pubblico, ma è universalmente nota per la fama che da oltre quattrocento anni la circonda, legata alle vicende della riforma Gregoriana del calendario.

Padre Egnazio Danti (1536-1586), domenicano fiorentino, già cosmografo del granduca di Toscana, fu chiamato a Roma da Bologna dal papa Gregorio XIII, Boncompagni, per partecipare ai lavori della commissione del calendario in cui lavorava anche il gesuita Cristoforo Clavio (1536-1612), e ricoprire poi la cattedra episcopale di Alatri.

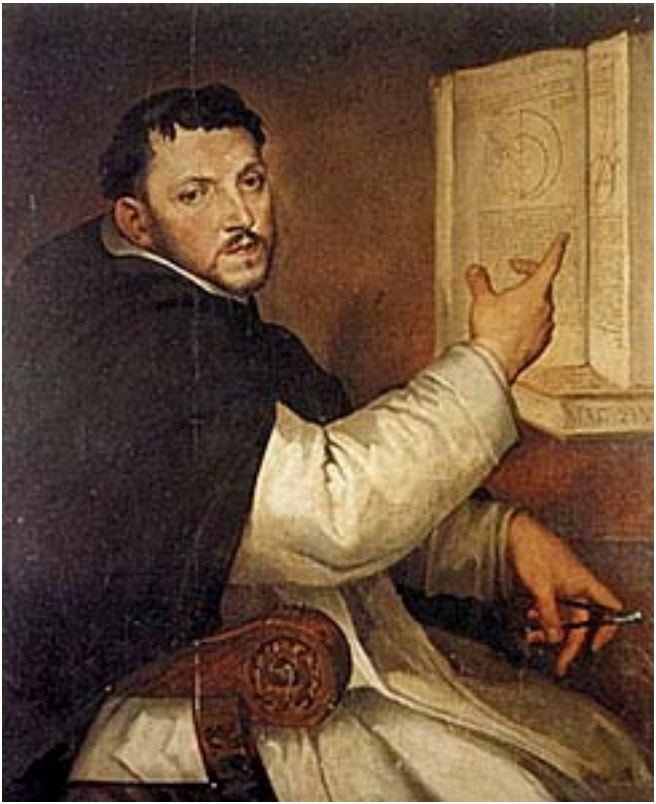

Fig. 1 Padre Egnazio Danti, domenicano.

Il Danti fu un personaggio chiave per l'astrometria solare del cinquecento poiché riportò in auge la tecnica delle meridiane a camera oscura appresa a Firenze, dove Paolo dal Pozzo Toscanelli (1397-1482) ne aveva realizzata una "per studiare il Sole in certi dì dell'anno"[1] il cui foro gnomonico si trovava proprio nella cupola del Brunelleschi.

Danti ebbe il merito di capire che questi strumenti potevano essere consultati pubblicamente, e ne avviò la costruzione in Santa Maria Novella sia fuori sulla facciata[2] di Leon Battista Alberti che dentro,[3] e a Bologna in San Petronio. Danti pubblicò diversi lavori, e tra essi la Tractatio Usi Gnomoni Magni (Bologna 1577), su un foglio solo, che mostra la semplicità ed al contempo la precisione raggiungibile con questo tipo di telescopi senza lenti per l'osservazione del Sole. Qualche anno dopo l'introduzione del telescopio sia Galileo (1613)[4] che Keplero mostreranno interesse verso questi strumenti perché capaci di mostrare anche le macchie solari più grandi.

La grande attività divulgativa serviva, secondo il Danti, a schiantare dai popoli la credulonerìa nei deliri dell'astrologia, e gli strumenti da lui realizzati restano a testimonianza di ciò.

## La meridiana del Danti nella Torre dei Venti

In linea con l'attività scientifico-divulgativa del Danti si colloca senz'altro la meridiana tracciata nella torre dei Venti, che la tradizione vuole sia stata visitata anche da Gregorio XIII, probabilmente il 21 marzo 1581 come suppone il padre Stein,[5] per convincersi dell'anticipo ormai arrivato a dieci giorni dell'equinozio di primavera sulla data che il concilio di Nicea aveva fissato al 21 marzo per il computo pasquale.

La Torre dei Venti voluta da Gregorio XIII fu costruita dall'architetto Ottaviano Mascherino, l'altezza raggiunge i 73 m ed è il punto più alto del Vaticano dopo la cupola di S. Pietro. Ad Atene nell'agorà romano esisteva la Torre dei Venti, chiamata anche *horologion*, una torre ottogonale in marmo Pentelico, ed a questo modello ci era ispirati nel dare il nome alla Torre.

La sala, splendidamente affrescata da Nicola Circignani detto il Pomarancio, su indicazioni del Danti stesso e presenta due finestre orientate ad Est e nel muro meridionale è praticato un foro conico, alloggiato in una nicchia esterna profonda circa 40 cm e chiusa da un vetro che impedisce alla pioggia e al vento di entrare. Da questo foro, chiamato gnomonico perché per suo tramite si conosce (dalla stessa radice del greco *gnosco*) il moto del Sole, entrano i raggi del Sole.

La sala è probabilmente le stessa dove ebbero luogo alcune discussioni della Commissione per il calendario, dato che essa è conosciuta anche col nome di sala del calendario.[6] Al centro del soffitto Danti aveva fatto dipingere una grande rosa dei venti in corrispondenza con l'indicatore dell'anemografo.

---

[1] F. Mazzucconi, P. Ranfagni, A. Righini, *Leonardo Ximenes S. J. e il grande gnomone di Santa Maria del Fiore, in Firenze*, in F. Bònoli et al. Il Sole nella Chiesa, Giornale di Astronomia **32**, p. 83-90, Bologna, 2006.

[2] S. Bartolini, *Gli strumenti astronomici di Egnazio Danti sulla facciata di S. Maria Novella*, Polistampa, Firenze, 2008.

[3] S. Bartolini, *I Fori Gnomonici di Egnazio Danti in Santa Maria Novella*, Polistampa, Firenze, 2006.

[4] G. Galilei, *Istoria e dimostrazioni sopra le macchie solari*, Roma 1613, p. 53-54.

[5] J. W Stein S. J., *The Meridian Room in the Vatican "Tower of the Winds"*, Miscellanea Astronomica vol III art. 97, Specola Vaticana, Città del Vaticano, 1950.

[6] Lo attesta un'iscrizione voluta dal Card. Zelada e composta dall'Abate Morcelli nel pontificato di Pio VI, posta sulla porta che dalla Biblioteca va alla Specola. Pio Vi, Pontefice Massimo, nell'anno XXIII del suo Pontificato, 1797, essendo Marino Carafa Prefetto dei SS. Palazzi Apostolici, dopo aver restaurato ed ampliato la torre osservatorio, dove matematici insigni tennero le loro sedute per la riforma gregoriana del calendario, aprì questo ingresso che permette di salire dalla Biblioteca, per affidare la sede (dell'osservatorio) e il luogo (che lo ospita) alla tutela del Cardinale

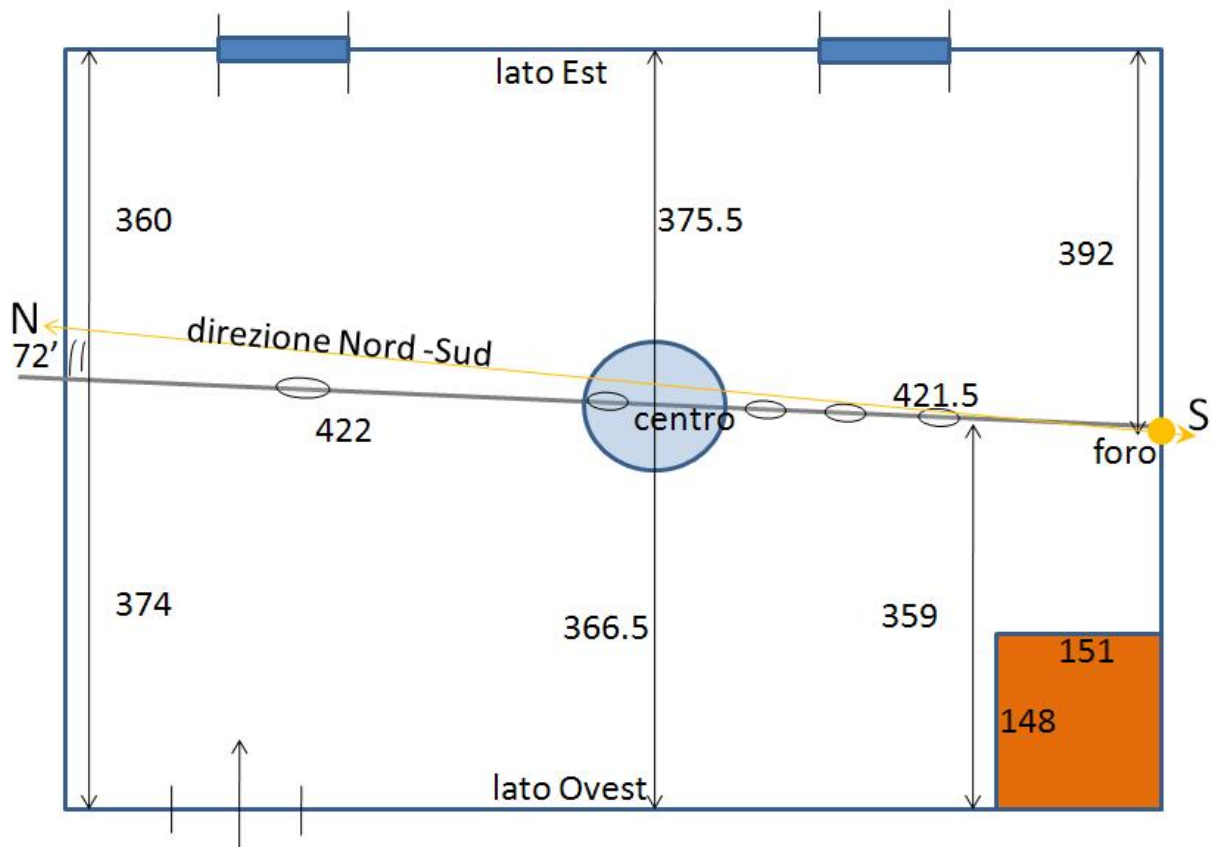

Fig. 2 Pianta della sala della meridiana della Torre dei Venti: l'orientamento generale segue solo approssimativamente i punti cardinali.

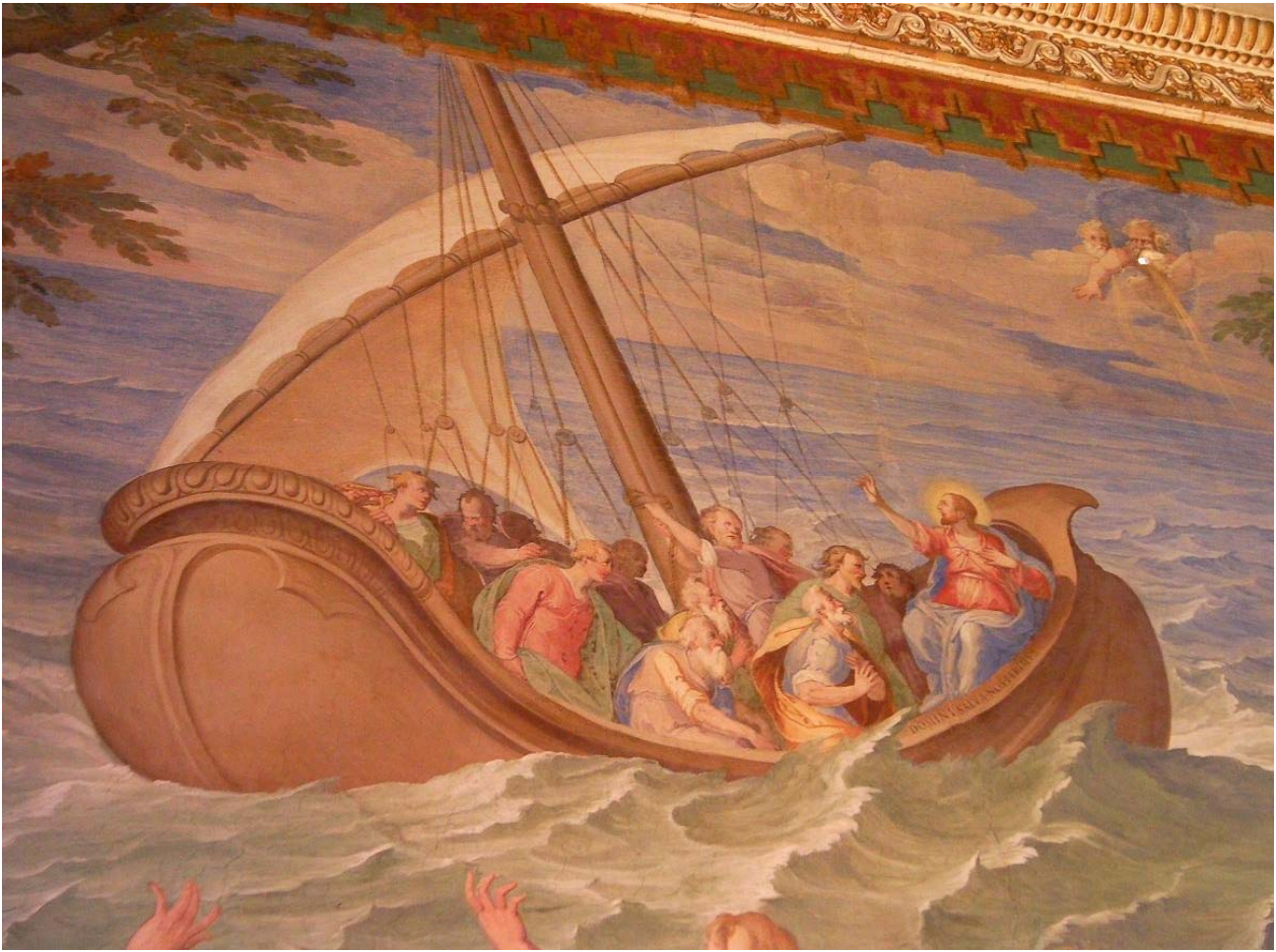

Fig. 3 Il foro gnomonico è nella bocca del genio del vento meridionale Austro, in alto a destra sopra la barca di Pietro nell'episodio della tempesta sedata [Matteo cap. **8**, 23-33].[7]

Sotto questo foro, chiamato tecnicamente foro stenopeico o foro gnomonico, a seconda che si voglia sottolineare che è un foro piccolo oppure che mediante il quale si "conosce" il moto annuale del Sole, parte una linea orientata verso Nord che l'immagine stenopeica del Sole attraversa ad ogni mezzogiorno locale. Il fenomeno è ben visibile anche con le finestre aperte, e dura per oltre due ore a cavallo del mezzogiorno locale. Esaminiamo di seguito alcune caratteristiche tecniche di questa celebre meridiana in seguito alla ricognizione astrometrica condotta nei mesi di febbraio e marzo 2009.

A queste aggiungiamo anche alcune considerazioni storiche, estese fino alla presenza del Mons. Filippo Luigi Gilii, che tracciò al piano superiore un'altra meridiana a foro stenopeico (1797) oggi non più funzionante perché il foro è stato chiuso, ed una murale sulla facciata orientale del terrazzo (1797 anch'essa).

---

[7] I passi biblici rappresentati negli affreschi della sala della Meridiana sono sul muro Sud: Matteo 8, 23-33 la tempesta sedata sul mare di Galilea; Genesi 41, 5-7 il sogno del Faraone; Genesi 1, 1-2 lo Spirito di Dio sulle acque; Esodo 10, 13-15 la piaga delle locuste; angolo Sud Ovest: Apocalisse 7, 1-4 i quattro venti; Esodo 10, 18-20 Mosè che comanda al vento di scacciare le locuste; Ovest: Atti degli Apostoli 27, 23-25 naufragio di S. Paolo a Malta; Atti 28, 1-5 S. Paolo e la vipera; Daniele 7, 2-3 i quattro venti e le quattro bestie; III Re, 18, 42-45 Elia prega sul monte Carmelo per la pioggia; Ezechiele 1, 1-5 visione di Ezechiele; muro Nord: Geremia 1, 14-15 Ab Aquilone pandetur omne malum; Matteo 7, 26-27 Parabola di colui che costruisce la casa sulla sabbia; Ecclesiaste (Qoelet) 11, 4 colui che osserva i venti; Giobbe 1, 13-15 e 18-19, un forte vento abbatte la casa dove mangiavano i figli di Giobbe; Giobbe 19, 23-24 lamento; muro Est Salmi 134, 7; Esodo 14,21 il vento d'oriente che soffiò durante la notte asciugando il Mar Rosso; Giona 1, 15 -2, 1 Giona è gettato in mare in bocca alla balena; Matteo 2,1-12 Adorazione dei Magi, da Stein op. cit. p. 38-41.

# Ricognizione astrometrica e riduzione dati

Si parla di ricognizione perché abbiamo misurato i parametri essenziali della meridiana: altezza del foro, posizione dei segni zodiacali a terra e deviazioni eventuali dalla rettilineità della linea stessa.

Si aggiunge il termine "astrometrica" perché questi parametri sono stati ricalcolati anche valendosi delle effemeridi del Sole confrontate con le misure prese in loco, in particolare l'altezza del foro è stata anche ricalcolata a partire dalla misura della posizione dell'immagine a terra e dell'angolo che il Sole formava in quel momento con l'orizzonte. Allo stesso modo è stata misurata la deviazione generale della linea dal Nord Celeste, dall'anticipo rispetto alle effemeridi, con cui il centro dell'immagine del Sole transita sulla linea meridiana. Questo è stato fatto con l'ausilio di un filmato realizzato a 60 fotogrammi per secondo.

L'altezza del centro del foro stenopeico risulta di 5 m 18 cm, sia da misure dirette che calcolate sulla posizione dell'immagine del Sole lungo la linea meridiana. Il dato è in ottimo accordo con quello pubblicato dal gesuita padre Stein (1950) di 5 m 19 cm.

Il foro ha un diametro di 14 mm ed è 2 cm rientrato rispetto al muro.[8]

Già Giuseppe Calandrelli nel 1842 aveva trovato la deviazione maggiore di 1 grado, ed il padre Francesco Denza, Barnabita e primo direttore della Specola Vaticana appena ripristinata da Leone XIII, nel 1892 aveva misurato tra il 21 marzo ed il primo aprile, una deviazione compresa tra 69' e 70'.[9]

---

[8] Questo dato sul foro è quello del rilievo originale del padre Stein (1950). Nella ricognizione astrometrica è la posizione del Sole calcolata con le effemeridi e la trigonometria a consentire di ricostruire le dimensioni dello strumento. Il diametro del foro si ottiene dal tempo di transito del Sole sulla linea che il 7 marzo 2009 valeva 145 s, inserito nella formula (1) con la latitudine del luogo (41.9°) e declinazione del momento (-5.1°) dà 49.34'; le effemeridi danno 32'.27. La semidifferenza 8'.53 è pari al diametro del foro visto dal piano della meridiana. La distanza foro-piano valeva 5.18 m/tan(43°) per cui il diametro del foro è pari a tan(8'.53)·5.18 m/tan(43°)=13.8 mm, compatibile con la misura fornita dal padre Stein.

[9] J. W. Stein S. J., op. cit., p. 36.

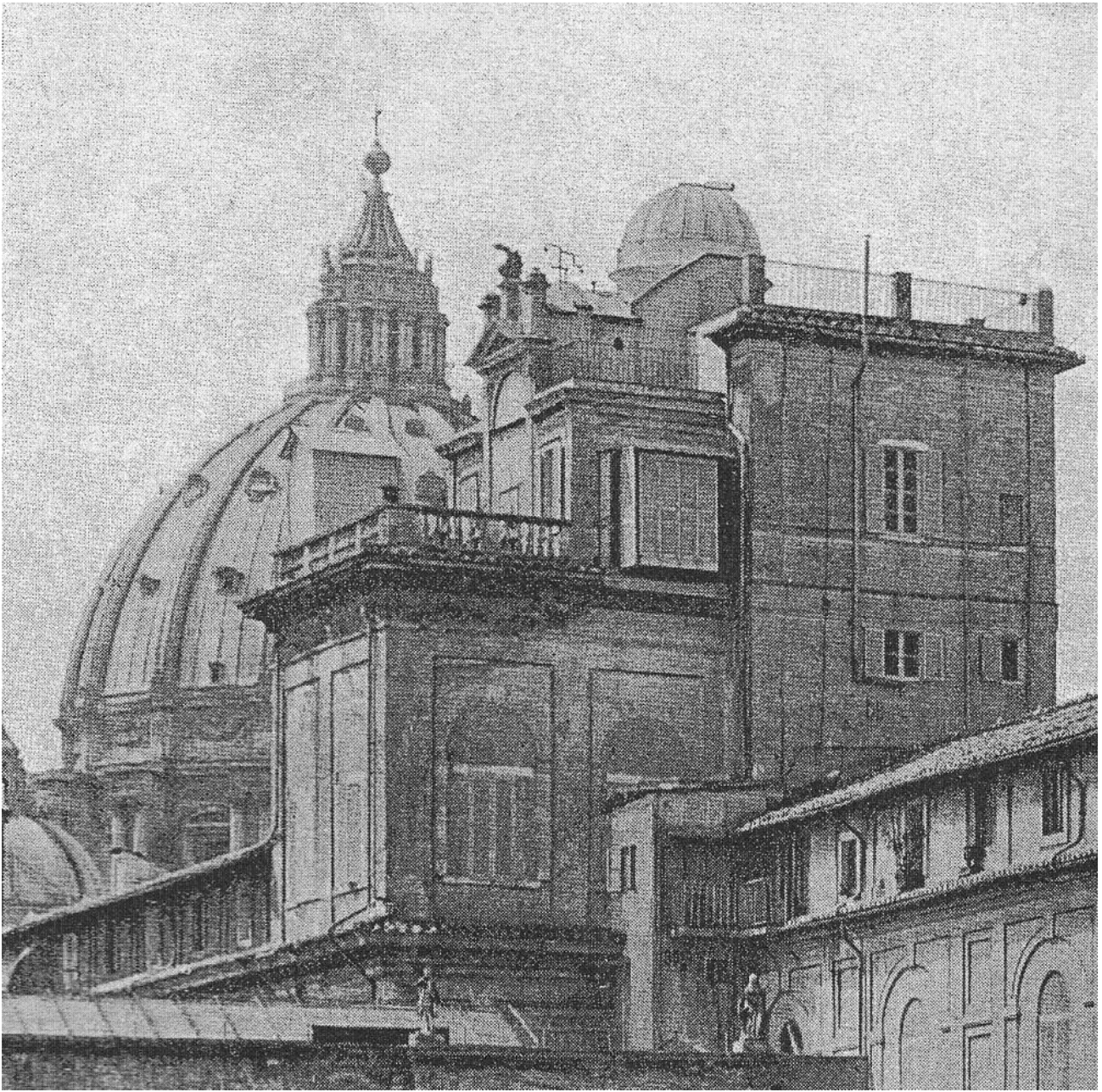

Fig. 4 Cupola della rifondata Specola Vaticana (1888) posta sopra la Torre dei Venti in una foto d'epoca.[10]

L'orientamento della linea meridiana è di circa 72' verso Ovest, così che l'immagine del Sole intercetta la linea con un anticipo di circa 4 minuti rispetto al mezzogiorno vero tra il 18 ed il 21 febbraio. I rilievi del padre Denza, ridotti dal padre Gesuita Stein hanno mostrato un disallineamento di circa 69' con un corrispondente anticipo di circa 3 minuti.

---

[10] Il Padre Denza propose al S. Padre Leone XIII di far collocare gli strumenti [esposti in occasione del suo giubileo sacerdotale avvenuto nel 1888] nella Torre dei Venti di Gregorio XIII, da tempo abbandonata; là, in quella che era stata, fino alla morte del Gilii (1821), la Specola Vaticana, si potevano riprendere le osservazioni che avrebbero portato nuova gloria all'antico luogo di studi e ricerche. […] Sopra la Torre dei Venti fu subito installata una cupola girevole di 3.5 metri e apertura di 58 cm, la prima delle quattro che, dopo qualche anno, sarebbero sorte in Vaticano, e sotto di essa fu collocato il piccolo equatoriale di Merz [di 10.2 cm di apertura]. Citazioni da S. Maffeo S.J., op. cit., p. 29-31.

La differenza nel tempo è normale e dipende dalla formula che mette in relazione la declinazione solare $\delta$ e la latitudine $\lambda$ del luogo con il ritardo o anticipo $\Delta t$ dovuto alla deviazione x (per il ritardo x è verso Est, mentre per l'anticipo x è verso Ovest) della linea rispetto al vero Nord Celeste.

Questa formula[11] è

(1) tan(x)=tan(15·$\Delta$t·cos$\delta$)/cos(90°-$\lambda$+$\delta$)

La differenza tra il valore di x=72' a febbraio 2009 e x=69' a marzo 1892 è dovuto all'uso di una formula più approssimata da parte del padre Stein (senza il cos$\delta$) nella riduzione dei dati. Ho utilizzato la formula più precisa ed ho ottenuto l'orientamento corrispondente ai vari giorni di misurazione.

| delta T [s] | giorno | Arcmin |
|---|---|---|
| 231.43 | 18/02/2009 | 70.62 |
| 235.78 | 19/02/2009 | 72.38 |
| 231.38 | 21/02/2009 | 71.88 |
| 217.27 | 07/03/2009 | 73.93 |
| 183 | 21/03/1892 | 70.09 |
| 181 | 22/03/1892 | 69.88 |
| 182 | 23/03/1892 | 70.83 |
| 179 | 31/03/1892 | 74.37 |
| 169 | 01/04/1892 | 70.81 |

Tabella 1. Anticipi delta T misurati in secondi del transito sulla linea rispetto alle effemeridi, e corrispondenti deviazioni della linea dal vero Nord.

Le misure in giallo sono prese col video, sincronizzato col tempo universale, durante il transito meridiano, e danno come valore 72.13'±0.25', includendo anche quelle rosa (estrapolate dalle foto prese dopo il transito meridiano il 18/2/09 ed il 7/3/09) abbiamo 72.2'±1.4', mentre le misure del padre Denza del 1892 danno 71.2'±1.8'.

Da queste considerazioni tecniche si nota che la misura presa con il video è compatibile entro gli errori sperimentali con quella presa ad occhio nudo dal padre Denza, che presenta una dispersione maggiore.

L'ampiezza di 1' a 7 metri dalla base del foro stenopeico corrisponde a 2 mm. Con una livella laser si è verificato che le deviazioni dalla retta sono comprese entro 3 mm lungo tutta la linea.

Le posizioni dei segni zodiacali, individuati tramite delle ellissi con i simboli zodiacali corrispondenti, non corrisponde esattamente alle posizioni che avrebbero dovuto avere nel 1580. Ci sono delle differenze anche oltre i 10 centimetri, tra le posizioni dei segni a terra e quelle che avrebbero dovuto avere secondo i calcoli. Il padre Stein fece misure simili e vi aggiunse anche la misura delle ellissi solari comparate con i valori calcolati, trovando varie discrepanze.

---

[11] La formula è approssimata meglio di 4 parti su 10000, il numero 15 è il rapporto tra 360°x3600" e 86400 s del giorno solare medio, ma il giorno solare vero, che è legato alla velocità angolare del Sole nel cielo, varia tra 86379 s all'equinozio di autunno e 86430 s a metà dicembre (C. Barbieri, *Lezioni di Astronomia*, Zanichelli, Bologna, 1999 p. 46).

| segni | distanze [cm] | da calcolo |
|---|---|---|
| pesci | 690.5 | 703.9 |
| equinozi | 471.5 | 464.8 |
| toro | 313.5 | 300.9 |
| leone | 207.5 | 204.6 |
| cancro | 159.75 | 172.3 |

Tabella 2. Distanze dalla base del foro stenopeico dei segni zodiacali, individuati dal centro dell'ellissi solare.

Infine la nicchia dove è alloggiato il foro non consente l'ingresso dei raggi solari per inclinazioni meridiane maggiori di circa 60 gradi, cioè qualche giorno dopo l'ingresso nel segno del Toro attorno al 23 aprile o prima dell'uscita dal Leone attorno al 21 agosto; in questo periodo la meridiana di fatto non funziona. A queste conclusioni si è giunti osservando che il cielo non è visibile dalla linea attraverso il foro già a 2 m 95 dalla base del foro stenopeico.

## L'errore dell'orientamento della linea verso il vero Nord

Nonostante la realizzazione di una meridiana perfettamente orientata sia stata un'operazione che non ha esentato da errori né Giandomenico Cassini (1655, meridiana di San Petronio a Bologna: 2' di deviazione verso Est) né Francesco Bianchini (1702, meridiana di Santa Maria degli Angeli a Roma: 4' 30" di deviazione verso Est), né prima di lui il Toscanelli (1475, meridiana di Santa Maria del Fiore a Firenze: 20' di deviazione verso Ovest) o Tycho Brahe (17' di deviazione verso Ovest, Uraniborg-Isola di Hven, Danimarca), il valore della deviazione dal Nord della meridiana del Danti nella Torre dei Venti è grande. Evidentemente il Danti non aveva intenzione di realizzare uno strumento di precisione.

Sono propenso piuttosto a ritenere che lo strumento sia solo dimostrativo, adattato e l'orientamento sia stato determinato dall'orientamento della stanza, e corredato dai segni zodiacali tratti da un modello concepito per latitudini leggermente maggiori di quella di Roma e copiato rapidamente per l'occasione.

Lo stesso fatto che il Sole non entra più ad illuminare la linea meridiana per ben quattro mesi a cavallo del solstizio estivo, e che non ci sia un prolungamento della linea verso il solstizio invernale (da metà novembre all'inizio di febbraio il Sole si muove sul muro settentrionale senza nessun riferimento meridiano sull'affresco) mostra che la linea è stata lavorata rapidamente, forse proprio per una visita del Papa in occasione dell'equinozio del 1581.

## La dimostrazione dell'arretramento della data dell'equinozio al 1581

Vale la pena spendere qualche parola sugli elementi necessari per completare la dimostrazione, mediante una sola visita alla meridiana, dell'errore che il calendario Giuliano aveva accumulato sull'equinozio astronomico.

Per costruire una meridiana a foro stenopeico è necessario conoscere oltre all'altezza del foro anche la latitudine del luogo. Questa infatti determina l'altezza sull'orizzonte del Polo Nord Celeste, ed il suo angolo complementare l'altezza massima dell'equatore celeste sul meridiano celeste Sud.

A Roma la latitudine è 41° 54', e l'equatore celeste culmina a 48° 06'.

Questo è anche l'angolo che il Sole equinoziale fa con l'orizzonte al momento della culminazione in meridiano. Con la trigonometria elementare si può ricavare che l'immagine stenopeica è proiettata sulla linea ad una distanza pari al 90% dell'altezza, che nel caso della meridiana della Torre dei Venti vale 4 m 66 cm.

Senza questa premessa trigonometrica anche il papa Gregorio XIII non avrebbe potuto apprezzare l'evidenza dei 10 giorni di anticipo con cui il Sole raggiungeva questa posizione, nonostante il calendario segnasse ancora la data dell'11 marzo.

L'ellissi solare corrispondente agli equinozi è centrata a 4 m 71 cm, appena 5 cm verso Nord rispetto al punto dove doveva trovarsi per la latitudine di Roma. Ai fini della dimostrazione dei giorni dell'equinozio quella meridiana ne anticipava l'identificazione di quasi un giorno, ma lo scopo dimostrativo era stato ottenuto.

Oggi il Sole continua a passare sull'ellissi equinoziale un giorno prima dell'equinozio vero per questo piccolo errore di posizionamento.

L'errore di posizionamento dell'ellissi corrispondente all'ingresso nel segno dei Pesci, data secondo effemeridi 18 febbraio 2009 h 14, data osservata 19 febbraio 2009 h 12, corrisponde ancora ad un intero giorno, circa 13 cm.

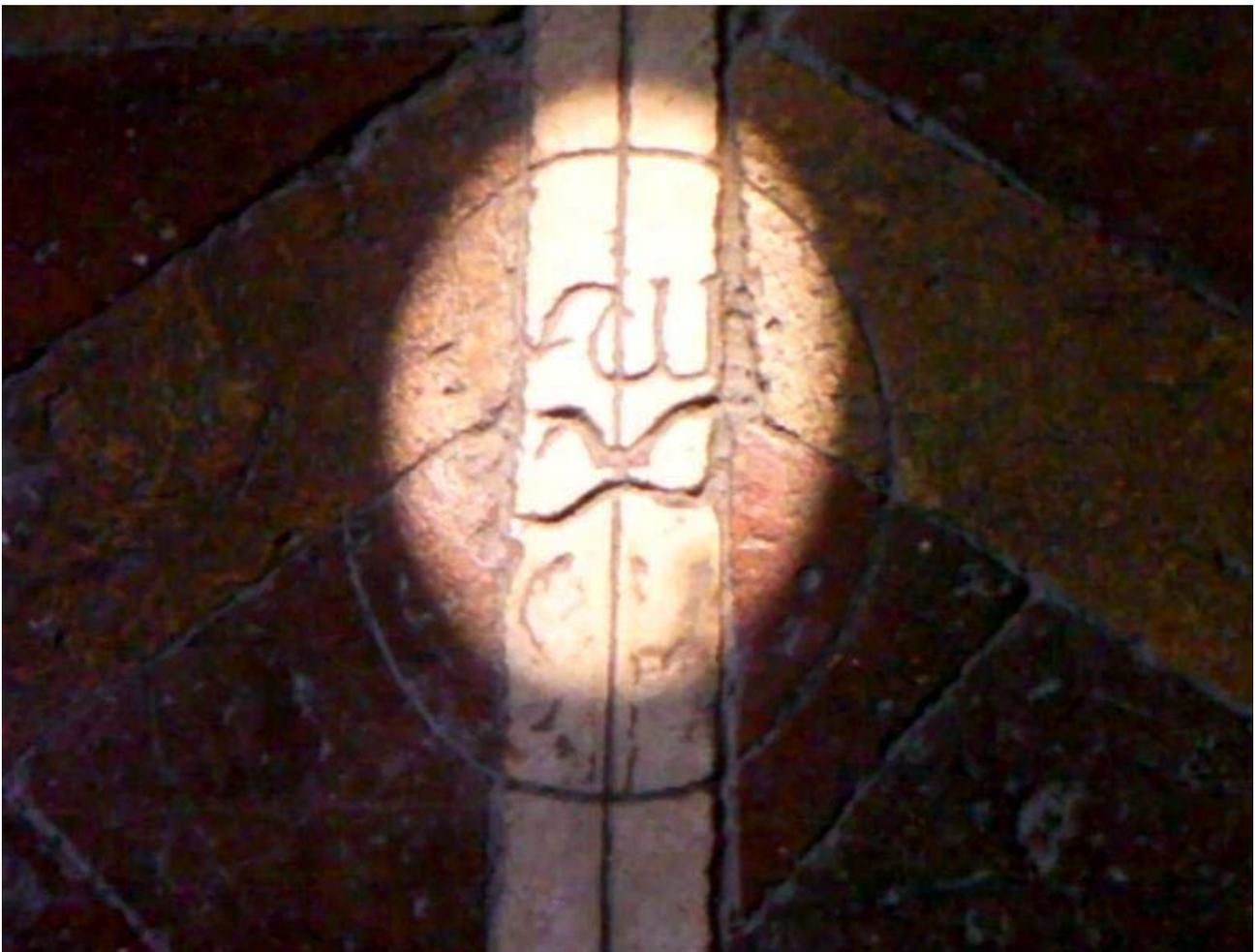

Fig. 5 Immagine del Sole al momento del transito sulla linea meridiana del 19 febbraio 2009. Il Sud è in alto.

Il Sole qui sembra entrato da qualche ora nel segno dei Pesci, invece ne sono già passate 23, ed in 23 ore il Sole il 19 febbraio percorre il 62% del suo diametro.

## Le meridiane del Gilii

Le due meridiane del Mons. Filippo Luigi Gilii (1797) al piano della Torre dei Venti che si trova sopra la sala della Meridiana, una interna a camera oscura ed una tracciata sul muro orientale della terrazza, testimoniano l'interesse per la gnomonica di questo monsignore, che si preoccupò in seguito di progettare e far tracciare la meridiana in piazza San Pietro nel 1817.[12]

Pietro Maccaranio prefetto della Reverenda Fabbrica di S. Pietro la sistemò per la pubblica utilità nel 1817, come è riportato nell'iscrizione posta alla base settentrionale dell'obeliso vaticano. Gilii attese a studi di gnomonica e meteorologia inaugurando un nuovo corso scientifico alla Torre dei Venti che dopo gli affreschi e la meridiana del Danti aveva visto solo l'augusta e fugace presenza della regina Cristina di Svezia (1626-1689).

Anche la meridiana del Gilii fu studiata dal padre Denza e la deviazione trovata tra 5'.4 (agli equinozi) e 6'.4 (al solstizio d'Inverno) verso Ovest.[13]  Oggi il foro stenopeico non è più aperto e quella meridiana non funziona più.

## Referenze:

---

[12] S. Maffeo S. J., op. cit., p. 7.
[13] J. W. Stein S. J., op. cit. 1950.